\documentclass{jpconf}

\usepackage{citesort}

\usepackage{epsfig}
\usepackage{graphicx}
\usepackage{amsmath}
\usepackage{amssymb}

\usepackage[small]{subfigure}

\newcommand{\ELYOs}{ELYO-$s$}
\newcommand{\ELYOsp}{ELYO-$s\!+\!p$~}
\newcommand{\rso}{r_s^\mathrm{o}}
\newcommand{\fs}{\footnotesize}

\begin{document}
\title{Lee-Yang-inspired functional including contributions to fourth order in effective field theory}
%
\author{J{\'e}r{\'e}my Bonnard}
\address{Universit{\'e} de Lyon, Institut de Physique des 2 Infinis de Lyon, IN2P3-CNRS-UCBL, 4 rue Enrico Fermi, 69622 Villeurbanne, France}
\ead{j.bonnard@ip2i.in2p3.fr}
\begin{abstract}
In the last years, a new family of energy density functionals directly inspired by effective field theory (EFT) has been developed, among which the ELYO (extended Lee-Yang Orsay) functional. In this paper, a new extension of ELYO that includes the recently calculated fourth-order EFT terms is presented.
Compared to the previous version, the description of neutron-drop energies slightly degrades for a harmonic trap $\hbar\omega =10$ MeV---yet remaining reasonable---, but improves for $\hbar\omega =5$ MeV. Furthermore, the obtained neutron effective mass agrees significantly better with \textit{ab-initio} estimates.
\end{abstract}
%

\section{Introduction}
%
In the last years, efforts have been undertaken to link energy density functionals (EDFs) to more fundamental approaches of nuclear systems, in particular effective field theory (EFT). Among the progress achieved so far, two novel EDFs characterized by singular density dependencies have been proposed, namely YGLO\footnote{Yang-Grasso-Lacroix Orsay.}~\cite{YGLO,drop,dropE} and ELYO\footnote{Extended Lee-Yang Orsay.}~\cite{drop,dropE,ELYOs,ELYOsp}. While both rely on the Lee-Yang formula~\cite{LY,LY_EEF}---re-derived more recently within the framework of EFT---to ensure a correct equation of state (EOS) of infinite nuclear matter at very low densities, they are based on different \textit{ansatzes} to describe density regimes of interest for nuclei. On the one hand, YGLO is a hybrid functional combining standard Skyrme terms with an innovative one whose form is borrowed from EFT resummation techniques. On the other hand, ELYO directly extends the validity domain of the Lee-Yang expansion through the definition of a density-dependent scattering length.

This paper focuses on the ELYO EDF. First designed by retaining only the contributions due to $s$-wave scattering~\cite{drop,dropE,ELYOs} (\ELYOs), it has then been generalized to also account for the $p$-wave terms~\cite{ELYOsp} (\ELYOsp), thus encompassing the full third-order Lee-Yang formula.
Here I propose to go one step further with the development of a new version of the functional that consists in including the fourth-order contributions to the EOS of dilute Fermi gas recently obtained in EFT~\cite{EFT4l,EFT4}. To this end, I follow step by step the same method than that in Ref.~\cite{ELYOsp} for the $p$-wave terms. First, the ELYO ansatz is enriched in Section~\ref{sec_PNM} for pure neutron matter (PNM). Next, in Section~\ref{sec_drops}, neutron drops are considered with the aim of rendering the EDF applicable to finite-size systems. In Section~\ref{sec_SNM}, the remaining adjustable parameters are constrained by symmetric nuclear matter (SNM) properties. Finally, conclusions are drawn in Section~\ref{sec_concl}.

\section{Equation of state of PNM based on the EFT expansion for the dilute Fermi gas} \label{sec_PNM}
The EOS of low-density PNM as obtained at fourth-order in EFT reads~\cite{EFT4l,EFT4}
\begin{equation}\label{LY4}
\begin{split}
   \dfrac{E}{N} = \dfrac{\hbar^2 k_F^2}{2m}\biggl\lbrace \dfrac{3}{5} & + \dfrac{2}{3\pi}(a_sk_F) + \dfrac{4}{35\pi^2} (11-2\ln2)(a_sk_F)^2  + \dfrac{1}{10\pi} (r_sk_F)(a_sk_F)^2  \\
  &  + \dfrac{1}{10}\gamma_1 (a_sk_F)^3+\dfrac{3}{5\pi} (a_pk_F)^3 + \gamma_2 (r_sk_F)(a_sk_F)^3 - \gamma_3  (a_sk_F)^4 \biggr\rbrace,
\end{split}
\end{equation}
with 
\begin{equation}
 \gamma_1 = 0.181853,  \qquad \gamma_2 = 0.0644872, \qquad  \gamma_3 = 0.0425,
\end{equation}
$m$ the nucleon mass, $k_F$ the Fermi momentum related to the density $\rho$ through $k_F=(3\pi^2\rho)^{1/3}$, $a_s = -18.9$ fm the neutron-neutron $s$-wave scattering length, $r_s = 2.75$ fm the associated effective range, and $a_p =0.63$ fm the $p$-wave scattering length (see discussion in Ref.~\cite{ELYOsp}).

The EOS \eqref{LY4}, which is an extension of the original Lee-Yang formula~\cite{LY,LY_EEF}, is valid
for $|a_sk_F| < 1$, that is $\rho \lesssim 5\times 10^{-6}$ fm$^{-3}$. Besides the new fourth-order terms (the two lasts), there is a difference with the expression employed in Ref.~\cite{ELYOsp} where the coefficient $\gamma_1$ was equal to 0.19. This small discrepancy ensues from the fact that, there, the third-order EFT expansion was coming from another source~\cite{LY_EEF}. Here is kept the value calculated together with the new fourth-order contributions for consistency; both being anyway compatible. 
It is worth noticing that the general expression published in Ref.~\cite{EFT4l,EFT4} contains a logarithmic component that cancels for PNM due to Pauli principle.

Let me now extend the ELYO functional by including fourth-order contributions. 
First, as for the former versions, the new ELYO EOS for PNM is constructed, starting from the expansion~\eqref{LY4}, by considering that the $s$-wave scattering length depends on the density $\rho$, in such a way as to broaden the validity condition up to any density scale, i.e. such that $|a_s(\rho)k_F| < 1$, $\forall \rho$. The ELYO ansatz for the form of this density dependence, which can be viewed as mean to model in-medium effects, is
\begin{gather} \label{as_rho}
a_s(\rho)=
\left\lbrace
\begin{array}{llc}
 a_s                          & \text{if } |a_sk_F| <    1 \\
- 1/(3\pi^2\rho)^{1/3}  & \text{if } |a_sk_F| \geq 1.
\end{array}\right.
\end{gather}
When no explicit dependence on the density is specified, $a_s$ stands for the physical value.

\begin{figure}[t!]
\begin{center}
\begin{minipage}[!b]{0.48\textwidth}
\includegraphics[width=\textwidth]{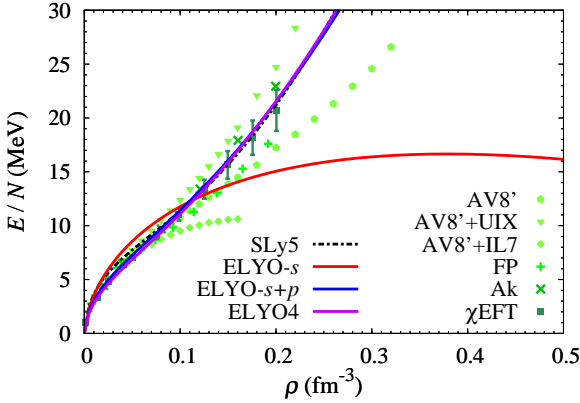}%
\end{minipage}\hspace{0.0\textwidth}
\begin{minipage}[!b]{0.45\textwidth}\vspace{-0.55cm}
\caption{EOS of PNM as obtained with the ELYO-$s$ (red), \ELYOsp~(blue), and ELYO4 (purple) EDFs. Also shown are the EOS for SLy5 (black dashed), and \textit{ab-initio} results (green) from Refs.~\cite{NCSMQMC,QMC} (AV8', pentagons; AV8'+UIX, triangles; AV8'+IL7, diamonds), Ref.~\cite{QMCFP} (FP, plus signs), Ref.~\cite{QMCAkmal} (Ak, crosses), and Ref.~\cite{DSS} (squares, with error bars).}
\label{fig_EoSPNM}
\end{minipage}
\end{center}
\vspace{-0.9cm}
\end{figure}

The question about how to treat the other physical constants $r_s$ and $a_p$ beyond the dilute regime has been discussed in detail in Ref.~\cite{ELYOsp}, and I adopt here the same approach: The effective range is adjusted and the tuned value is used at any density scale, while the $p$-wave scattering length keeps its physical value even in the region where $a_s$ evolves with $\rho$.  As benchmark for the fit of $r_s$, the PNM EOS of the SLy5 Skyrme parameterization~\cite{Sly5}, which well agrees with available \textit{ab-initio} estimates, is considered. The optimal value of the tunable effective range is denoted $\rso$, and the fit leads to $\rso = 6.831$ fm.  The corresponding PNM EOS of the new functional, named ELYO4 from now on, is displayed in Fig.~\ref{fig_EoSPNM}, where it is compared to the previous versions of ELYO, the SLy5 EDF, and a collection of \textit{ab-initio} results. As expected, the new EOS matches the one from \ELYOsp~that was already in good agreement with SLy5.
%

\section{Towards finite-size systems: Mapping to a Skyrme-like EDF} \label{sec_drops}
In order to completely define the ELYO4 EDF, the strategy consists in mapping the PNM EOS on a Skyrme-like one, that is by performing a term-by-term identification according to the power of $\rho$ occurring in Eq.~\eqref{LY4}, which allows to express the standard Skyrme parameters $\{t_i,x_i\}$ in function of the coefficients in the EFT expansion. Compared to the previous versions of ELYO, the new fourth-order contributions demand to consider an additional density-dependent Skyrme-like term $(t_{3^*},x_{3^*})$, associated with a power $\alpha^*=1$. Also, let me introduce as of now two extra parameters, $W_1$ and $W_2$, that modulate the contributions to the EOS from the Skyrme velocity-dependent terms, $(t_1,x_1)$ and $(t_2,x_2)$, respectively. Actually, a component proportional to $\rho^{5/3}$ in the EOS may be generated by such terms, by density-dependent terms with power $2/3$, or by any combination of these. The role of $W_1$ and $W_2$ is therefore to split the $\rho^{5/3}$ term---leaving the EOS unchanged---, and to act as weights to unambiguously determine each contribution \textit{via} an adjustment on the properties of finite-size systems. These two parameters turns out to be crucial since they correct the effective masses (see Refs.~\cite{ELYOsp,drop} for more detailed discussions).
Finally, the mapping reads:
\begin{subequations} \label{mapping}
\begin{align}
& t_0 \bigl(1-x_0(\rho)\bigr)  = \dfrac{4\pi\hbar^2}{m} a_s(\rho) , &
& t_3 \bigl(1-x_3(\rho)\bigr)  = \dfrac{144\hbar^2c_0}{35m}(11-2\ln 2) a_s^2(\rho) ,\\
& t_1 \bigl(1-x_1(\rho)\bigr)  = W_1 \dfrac{2\pi\hbar^2}{m} B_s(\rho) , &
& t_{3'} \bigl(1-x_{3'}(\rho)\bigr) = (1-W_1) \dfrac{18\pi\hbar^2 c_0^2}{5m} B_s(\rho), \label{t3p} \\ 
& t_2 \bigl(1 + x_2\bigr) = W_2 \dfrac{4\pi\hbar^2}{m} a_p^3 ,&
& t_{3''} \bigl(1-x_{3''}\bigr) = (1-W_2) \dfrac{108\pi\hbar^2c_0^2}{5m}  a_p^3, \label{t3pp} \\ 
& t_{3^*} \bigl(1-x_{3^*}(\rho)\bigr) = \dfrac{108\pi^4\hbar^2}{m} C_s(\rho),
\end{align}
\end{subequations}
with $c_0 = (3\pi^2)^{1/3}$, $\alpha = 1/3$, $\alpha'=\alpha''=2/3$, $\alpha^* =1$, and
\begin{equation}
B_s(\rho) = \rso a_s^2(\rho) + \gamma_1\pi a_s^3(\rho), \qquad  C_s(\rho) = \gamma_2 \rso a_s^3(\rho) - \gamma_3 a_s^4(\rho) .
\end{equation}
Above, as made explicit by the notation, it is assumed that all the dependencies on the density from $a_s(\rho)$ are contained within the $x_i$ parameters, while the $t_i$'s are kept constant and will be adjusted on the SNM EOS below.
By construction, writing down the Skyrme EOS for PNM with these parameters leads to Eq.~\eqref{LY4} with $a_s$ and $r_s$ replaced with $a_s(\rho)$ and $\rso$, respectively, the terms involving the splitting coefficient $W_1$ ($W_2$) recombining to give back the $s$-wave ($p$-wave) component proportional to $\rho^{5/3}$. It is worth noticing that the $3'$ and $3''$ terms, corresponding to the same power of density, can be gathered into a unique one: $\bar{\alpha} = 2/3$, $t_{\bar{3}} = t_{3'} + t_{3''}$, and $t_{\bar{3}} (1-x_{\bar{3}}(\rho))$ given by summing the equations on the right sides of lines~\eqref{t3p} and~\eqref{t3pp}.
\begin{figure}[b!]
\begin{center}
\includegraphics[width=0.48\columnwidth]{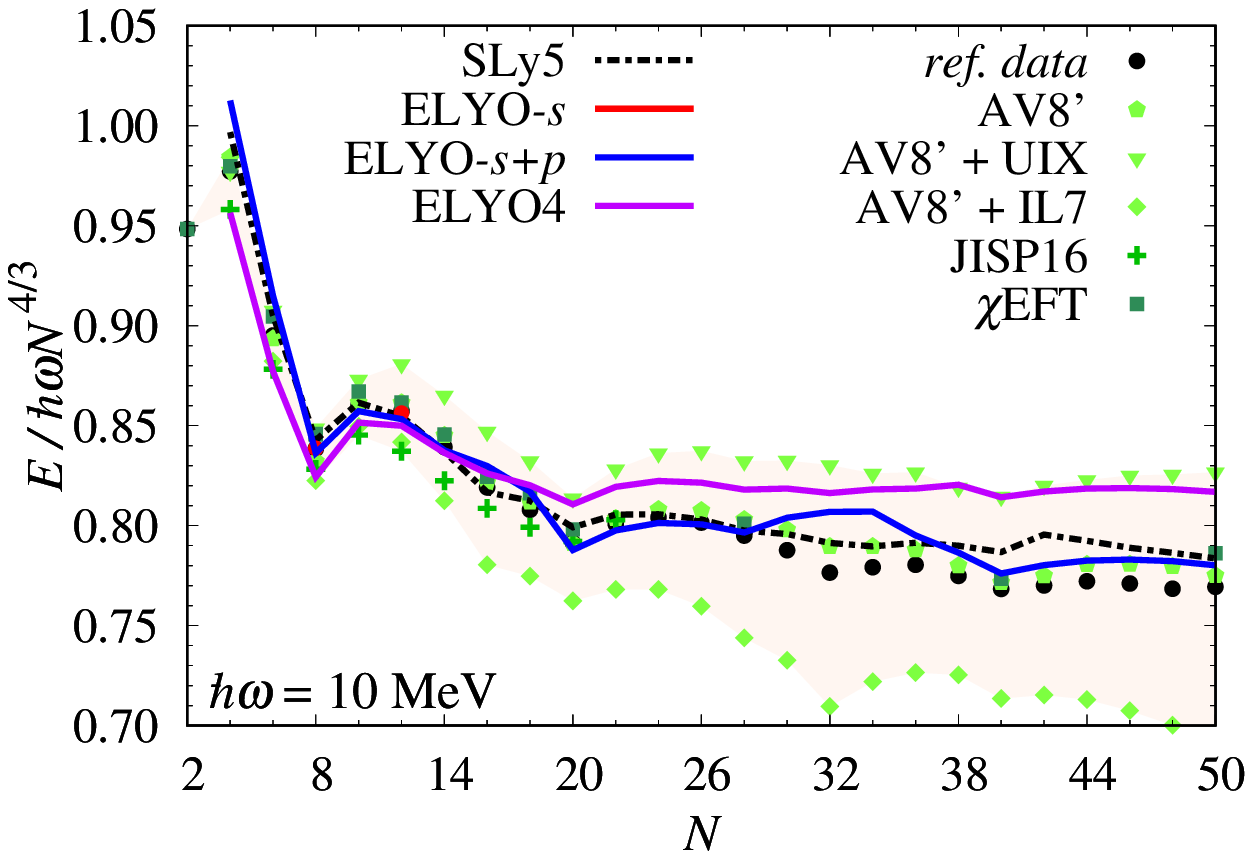}\hspace*{-0.18cm}\includegraphics[width=0.48\columnwidth]{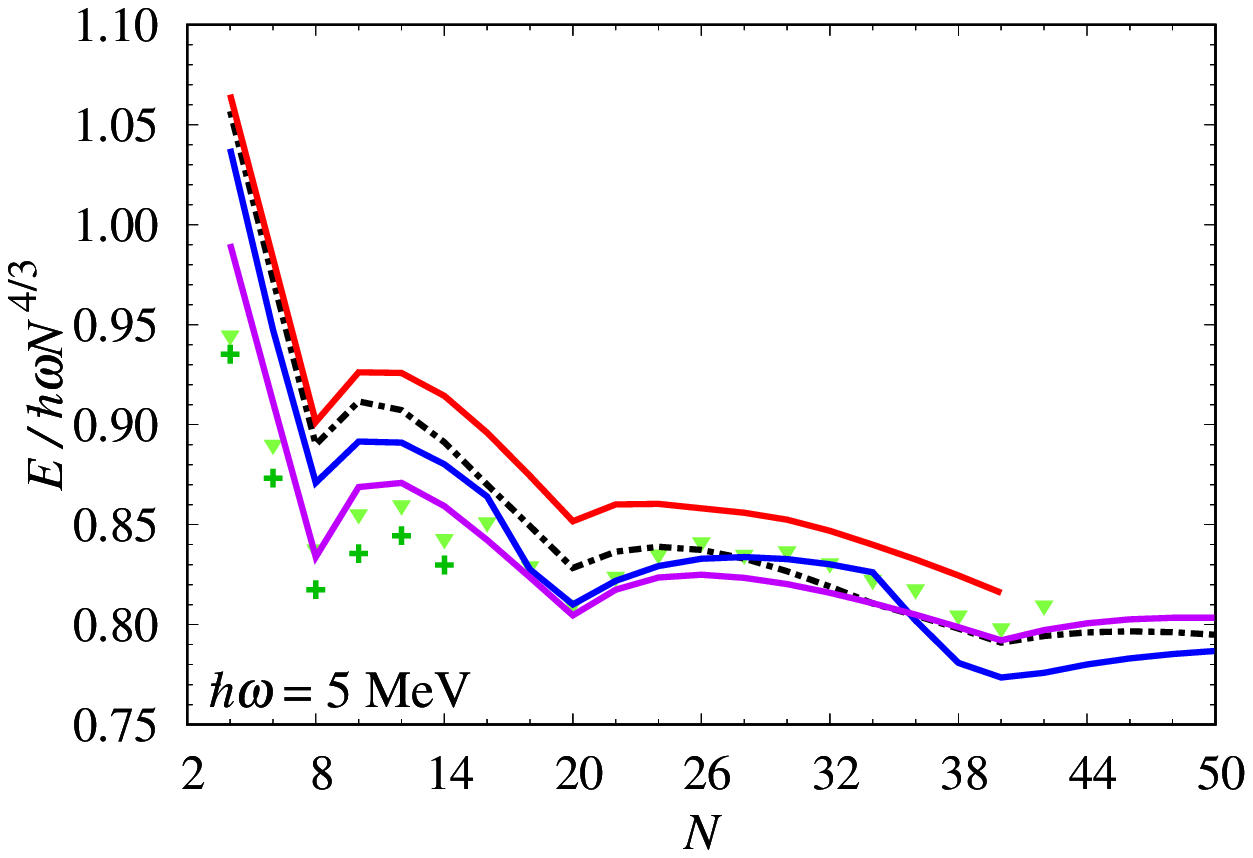}
\vspace*{-0.5cm}
\caption{Energies of neutron drops in traps of frequencies $\hbar\omega = 10$ (left) and 5 (right) MeV, scaled by $\hbar\omega N^{4/3}$, as obtained with the new ELYO4 (purple) and compared to the curves from the ELYO-$s$ (red), \ELYOsp~(bleu), and SLy5 (black dotted) EDFs. Also shown are \textit{ab-initio} estimates (green) from Refs.~\cite{NCSMQMC,QMC} (AV8', pentagons; AV8'+UIX, triangles; AV8'+IL7, diamonds), Ref.~\cite{NCSMQMC} (JISP16, plus), and Ref.~\cite{NCSMEFT} (squares). Their average, used as pseudo-data for the fit, is denoted as ``\textit{ref. data}" (black circles).}
\label{fig_EoSDrop}
\end{center}
\end{figure}

The advantage of such a mapping lies in the possibility to easily deduce the general form of the EDF. Indeed, substituting in the usual Skyrme energy density the  $(t_i, x_i)$ parameters with those defined above (and the number density $\rho$ by the local density $\rho(\vec{r})$ at the coordinate $\vec{r}$) directly yields the expression of the ELYO4 functional, which allows a direct implementation to finite-size systems. However, since these parameters have been designed from PNM, I shall exclusively focus in this work on finite systems composed of neutrons only, i.e. neutron droplets trapped in an isotropic harmonic field of frequency $\omega$. 
Considering even numbers of neutrons in spherical symmetry, the energy density then reduces to a functional involving only the following four local densities (see Ref.~\cite{HFBJ} for their formal definitions): $\rho(\vec{r})$, the kinetic density $\tau(\vec{r})$, the spin-current density $J(\vec{r})$, and the anomalous pairing density $\tilde{\rho}(\vec{r})$ (below, the index $n$ and $p$ respectively refer to their counterparts for neutrons and protons only). The central part of the energy density may eventually be cast as
\begin{equation} \label{ELYO4} 
\begin{split}  
\mathcal{E}_c = \mathcal{E}^\mathrm{Sk}_c &- \biggl[X_0 a_s[\rho] + X_3 \rho^{\alpha} a_s^2[\rho] + X_{3^*} \rho^{\alpha^*}C_s[\rho] + X_{\bar{3}} \rho^{\bar{\alpha}}D_s[\rho] \biggr] \biggl[\dfrac{1}{2}\rho^2  - \sum_{q=n,p}\!\!\rho_q^2 \biggr]  \\
&-W_1X_1 B_s[\rho] \biggl[\dfrac{1}{2}\rho\tau+\dfrac{3}{8}(\vec{\nabla}\rho)^2 -\dfrac{1}{4}\vec{J}^{\,2} -\sum_{q=n,p}\!\!\Big(\rho_q\tau_q+\dfrac{3}{4}(\vec{\nabla}\rho_q)^2\Bigr) \biggr]\\ 
& +  W_2X_2  a_p^3 \;\quad\biggl[\dfrac{1}{2}\rho\tau-\dfrac{1}{8}(\vec{\nabla}\rho)^2 - \dfrac{1}{4}\vec{J}^{\,2} + \sum_{q=n,p}\!\!\Big(\rho_q\tau_q-\dfrac{1}{4} (\vec{\nabla}\rho_q)^2\Bigr) \biggr], 
\end{split}
\end{equation}
in which $D_s[\rho] = \frac{1}{2}(1-W_1) B_s[\rho] + 3(1-W_2) a_p^3$, and
\begin{align}
& X_0         = \dfrac{2\pi\hbar^2}{m}               , &
& X_1         = \dfrac{\pi\hbar^2}{2m}               , &
& X_2         = \dfrac{\pi\hbar^2}{m}                , \\
& X_3         = \dfrac{12c_0\hbar^2}{35m}(11-2\ln 2) , &
& X_{3^*} \!\!= \dfrac{9\pi^4\hbar^2}{m}             , &
& X_{\bar{3}} = \dfrac{3\pi c_0^2\hbar^2}{5m}.
\end{align}
$\mathcal{E}^\mathrm{Sk}_c$ stands for the central part of a Skyrme EDF with three density-dependent terms such as $x_i^\mathrm{Sk} =1$ ($-1$) for $i=0,1,3,\bar{3}, 3^*$ (2), and that conveniently cancels out for neutron droplets.

\begin{figure}[t!]
\begin{center}
\begin{minipage}[!b]{0.48\textwidth}
\includegraphics[width=\textwidth]{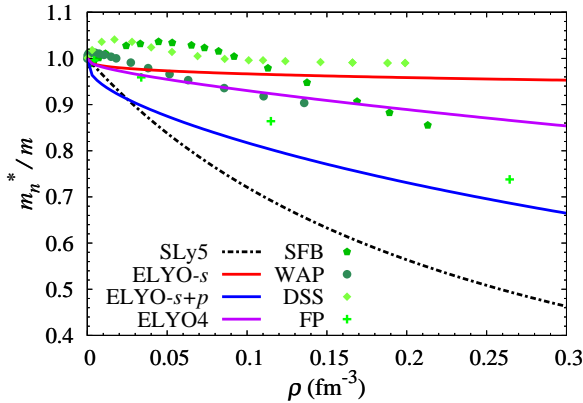}%
\end{minipage}\hspace{0.0\textwidth}
\begin{minipage}[!b]{0.45\textwidth}\vspace{-0.55cm}
\caption{Neutron effective masses $m_n^*/m$ versus the density for the ELYO-$s$ (red), \ELYOsp~(blue), and ELYO4 (purple) EDFs, compared to the one of SLy5 (black dashed) and to \textit{ab-initio} estimates (green) extracted from Ref.~\cite{SFB} (SFB, pentagons), Ref.~\cite{WAP} (WAP, circles), Ref.~\cite{DSS} (DSS, diamonds), and Ref.~\cite{QMCFP} (FP, squares).}
\label{fig_mstarPNM}
\end{minipage}
\end{center}
\vspace{-0.9cm}
\end{figure}

Following Ref.~\cite{ELYOsp}, the new parameters $W_1$ and $W_2$, together with the spin-orbit coupling constant $V_\mathrm{so}$ and the strength $V_\mathrm{pp}$ of a mixed surface-volume pairing force, are adjusted to reproduce the average of a collection of \textit{ab-initio} results for light drops $N\leq 20$ with $\hbar\omega=10$ MeV. The obtained parameters are reported in Table~\ref{tab_param}. For the pairing strength, the fit gives the minimum value allowed in the protocol, which corresponds to a drastic decrease compared to \ELYOsp$\!\!$.
The energies of neutron droplets for $\hbar\omega=5$ and 10 MeV are shown in Fig.~\ref{fig_EoSDrop}. Although for a 10-MeV trap the obtained values still live in the area of available \textit{ab-initio} estimates, a degradation in the description of the neutron-drop energies may globally be observed. In contrast, a slight improvement is to be noticed for $\hbar\omega=5$ MeV. In addition, the neutron effective mass comes now in better agreement with \textit{ab-initio} calculations, see Fig.~\ref{fig_mstarPNM}.

\section{Parameters from symmetric matter} \label{sec_SNM}
It remains now to determine the $t_i$ parameters to completely characterize the ELYO4 functional. First, to ensure that the previous introduction of $W_{1,2}$ do not modify the SNM EOS either, $t_{3'}$ and $t_{3''}$, which summed up give $t_{\bar{3}}$, have to be related to $t_1$ and $t_2$ through Eq.~(14) of Ref.~\cite{ELYOsp}. A fit on the SLy5 EOS of SNM then provides the values of $t_0$, $t_3$, $t_{3^*}$, and $\Theta_s = 3t_1+t_2(5+4x_2) = 3t_1+t_2 + \frac{16\pi\hbar^2}{m}a_p^3$.

At this stage, $t_1$ and $t_2$ are still not defined individually since $\Theta_s$ gives access only to the combination $3t_1+t_2$. Therefore, an additional equation is required. Contrary to what was claimed in Ref.~\cite{ELYOsp}, however, it is not necessary to resort to constraints on nuclei. Indeed, whereas by definition the splitting coefficients $W_{1,2}$ do not contribute to the EOS because the associated terms recombine, they do to the effective masses. This means that the isoscalar effective mass, which actually reads\footnote{The presence of the splitting parameters in $m_s^*$ has been wrongly omitted in Ref.~\cite{ELYOsp}, hence the value reported there for $m_s^*/m$ is not correct. Nonetheless, this has no consequence for the rest of the work.} $m_s^*/m = [1 + \frac{m}{8\hbar^2}\Theta^{\scriptscriptstyle W}_s \rho]^{-1}$ where $\Theta^{\scriptscriptstyle W}_s=3W_1t_1+W_2t_2 + W_2\frac{16\pi\hbar^2}{m}a_p^3$, offers the needed extra constraint. Imposing $m_s^*/m = 0.7$ at saturation density---a value close to the one of SLy5---, finally allows to finish the adjustment of the ELYO4 EDF, and
the parameters are summarized in Table~\ref{tab_param}. As a complement to Ref.~\cite{ELYOsp}, those of \ELYOsp are also reported, with $t_1$ and $t_2$ computed by applying the aforementioned constraint on $m_s^*$.

Table~\ref{tab_prop} shows some saturation properties of SNM as computed with the ELYO4 functional: the saturation density $\rho_c$, the energy per nucleon $E/A|_c$, the incompressibility $K_c$, the isoscalar effective mass; as well as the symmetry energy $J$, its slope $L$, and curvature $K_{\mathrm{sym}}$ at the parabolic approximation.
It can be seen that correct values are obtained, in particular $J$ and $L$ that both fall within the allowed ranges defined experimentally from heavy-ion collision~\cite{JLionalpha} and measurements of electric dipole polarizability on ${}^{208}$Pb~\cite{JLionalpha,JLalpha}, or conjectured from the unitary gas EOS~\cite{JLUGEOS}.
\begin{table}[t!]
\caption{Parameters of the \ELYOsp and ELYO4 functionals. The power in the density-dependent terms are $\alpha = 1/3$, $\bar{\alpha}=2/3$, and $\alpha^*=1$. $t_{\bar{3}}$ is  not adjusted (see text). $t_0$ and $V_\mathrm{pp}$ are given in MeV.fm$^3$;  $t_1$, $t_2$, and $V_\mathrm{so}$ in MeV.fm$^5$;  $t_3$ in (MeV.fm$^4$);  $t_{3^*}$ in MeV.fm$^6$.}
\begin{tabular*}{\linewidth}{@{\extracolsep{\fill}}lccccccccc}
\hline
\hline
        &       $W_1$  &  $W_2$       &     $t_0$      &     $t_1$      &     $t_2$      &     $t_3$      &   $t_{3^*}$    & $V_\mathrm{so}$  &  $V_\mathrm{pp}$  \\
\ELYOsp & {\fs -0.1760}& {\fs 0.4990} & {\fs -1916.91} & {\fs  -589.13} & {\fs  647.17 } & {\fs 15344.70} & {\fs     $-$ } & {\fs  81.20}     &  {\fs -252.14 }   \\
ELYO4   & {\fs 0.0675} & {\fs 0.1717} & {\fs -2270.61} & {\fs -6915.23} & {\fs 12750.98} & {\fs 28087.30} & {\fs 15843.90} & {\fs 103.32}     &  {\fs -150.00}    \\
\hline
\hline
\end{tabular*}
\label{tab_param}
\vspace*{-0.3cm}
\caption{Saturation properties of infinite matter as given by the ELYO4 functional.}
\begin{tabular*}{\linewidth}{@{\extracolsep{\fill}}ccccccccccc}
\hline
\hline
$\rho_c$ (fm$^{-3}$) & $E/A|_c$ (MeV) &    $K_c$ (MeV) & $m_s^*/m|_c$   &   $J$ (MeV)    &   $L$ (MeV)   &  $K_{\mathrm{sym}}$ (MeV)  \\
      0.162          &  $-$15.979       &      229.499   &    0.700       &    33.142      &    52.625     &  $-$113.930 \\
\hline
\hline
\end{tabular*}
\label{tab_prop}
\end{table}
 
\section{Conclusions} \label{sec_concl}
In this work I proposed an update of the ELYO EDF that consists in incorporating the fourth-order terms recently calculated in EFT. The new ELYO4 functional performs globally as well as the previous version \ELYOsp$\!\!$, the comparison with \textit{ab-initio} results degrading in some cases (energies of drops with $\hbar\omega=10$ MeV) and improving in others (energies of drops with $\hbar\omega=5$ MeV, neutron effective mass).

This new step in the development of ELYO highlights two points that deserves to be discussed and studied in more detail in the future. First, the price to pay for including new contributions in the functional is an increase of the number of adjustable parameters, which runs counter to the spirit of this family of EFT-inspired EDFs built with the aim of being less empirical than traditional ones. 
The fourth-order terms demands one more free parameter ($t_{3^*}$), bringing their total number to 10, i.e. exactly like for Skyrme EDFs. Consequently, this reduces the interest of EFT-inspired functionals. In addition, the perturbative nature of the low-density EFT expansion being lost at the level of the ELYO functional, there is no criterion to rely on to know when to stop including new terms.

The second point to consider concerns the finite-size instabilities. It has been pointed out in Ref.~\cite{ELYOsp} that \ELYOsp could be susceptible to present such instabilities. Taking into account the fourth-order terms could have solved the problem or, at the opposite, worsened it even if no convergence issue has been encountered in the present work. This question will be addressed in a forthcoming study based on linear response theory.

I  gratefully thank Marcella Grasso for fruitful discussions.

\section*{References}
\bibliography{paper_ELYO4}

\providecommand{\newblock}{}
\begin{thebibliography}{10}
\expandafter\ifx\csname url\endcsname\relax
  \def\url#1{{\tt #1}}\fi
\expandafter\ifx\csname urlprefix\endcsname\relax\def\urlprefix{URL }\fi
\providecommand{\eprint}[2][]{\url{#2}}

\bibitem{YGLO}
Yang C~J, Grasso M and Lacroix D 2016 {\em Phys. Rev.\/} C {\bf 94}(3) 031301

\bibitem{drop}
Bonnard J, Grasso M and Lacroix D 2018 {\em Phys. Rev.\/} C {\bf 98} 034319

\bibitem{dropE}
Bonnard J, Grasso M and Lacroix D 2021 {\em Phys. Rev.\/} C {\bf 103} 039901

\bibitem{ELYOs}
Grasso M, Lacroix D and Yang C~J 2017 {\em Phys. Rev.\/} C {\bf 95} 054327

\bibitem{ELYOsp}
Bonnard J, Grasso M and Lacroix D 2020 {\em Phys. Rev.\/} C {\bf 101} 064319

\bibitem{LY}
Lee T~D and Yang C~N 1957 {\em Phys. Rev.\/} {\bf 105} 1119--1120

\bibitem{LY_EEF}
Hammer H~W and Furnstahl R 2000 {\em Nucl. Phys.\/} A {\bf 678} 277--294

\bibitem{EFT4l}
Wellenhofer C, Drischler C and Schwenk A 2020 {\em Phys. Lett.\/} B {\bf 802}
  135247

\bibitem{EFT4}
Wellenhofer C, Drischler C and Schwenk A 2021 {\em Phys. Rev.\/} C {\bf 104}
  014003

\bibitem{Sly5}
Chabanat E, Bonche P, Haensel P, Meyer J and Schaeffer R 1997 {\em Nucl.
  Phys.\/} A {\bf 627} 710--746

\bibitem{NCSMQMC}
Maris P, Vary J~P, Gandolfi S, Carlson J and Pieper S~C 2013 {\em Phys. Rev.\/}
  C {\bf 87} 054318

\bibitem{QMC}
Gandolfi S, Carlson J and Pieper S~C 2011 {\em Phys. Rev. Lett.\/} {\bf 106}
  012501

\bibitem{QMCFP}
Friedman B and Pandharipande V 1981 {\em Nucl. Phys.\/} A {\bf 361} 502--520

\bibitem{QMCAkmal}
Akmal A, Pandharipande V~R and Ravenhall D~G 1998 {\em Phys. Rev.\/} C {\bf 58}
  1804--1828

\bibitem{DSS}
Drischler C, Som\`a V and Schwenk A 2014 {\em Phys. Rev.\/} C {\bf 89} 025806

\bibitem{NCSMEFT}
Potter H, Fischer S, Maris P, Vary J, Binder S, Calci A, Langhammer J and Roth
  R 2014 {\em Phys. Lett.\/} B {\bf 739} 445--450

\bibitem{HFBJ}
Dobaczewski J, Flocard H and Treiner J 1984 {\em Nucl. Phys.\/} A {\bf 422}
  103--139

\bibitem{SFB}
Schwenk A, Friman B and Brown G~E 2003 {\em Nucl. Phys.\/} A {\bf 713} 191--216

\bibitem{WAP}
Wambach J, Ainsworth T and Pines D 1993 {\em Nucl. Phys.\/} A {\bf 555}
  128--150

\bibitem{JLionalpha}
Lattimer J~M and Steiner A~W 2014 {\em Eur. Phys. J.\/} A {\bf 50} 40

\bibitem{JLalpha}
Roca-Maza X, Vi\~nas X, Centelles M, Agrawal B~K, Col\`o G, Paar N, Piekarewicz
  J and Vretenar D 2015 {\em Phys. Rev.\/} C {\bf 92} 064304

\bibitem{JLUGEOS}
Tews I, Lattimer J~M, Ohnishi A and Kolomeitsev E~E 2017 {\em Astrophys. J.\/}
  {\bf 848} 105

\end{thebibliography}
%
\end{document}